\RequirePackage{ifpdf}
\documentclass{PoS}
\usepackage{amsmath}

\title{Spectral functions of charmonium with 2+1 flavours of dynamical quarks}
\ShortTitle{Spectral functions of charmonium with 2+1 flavours of dynamical quarks}

\author{Szabolcs Bors\'anyi$^a$,
Stephan D\"urr$^{ab}$,
Zolt\'an Fodor$^{abc}$,
Christian Hoelbling$^{a}$,
S\'andor D. Katz$^{cd}$,
Stefan Krieg$^{ab}$,
Simon Mages$^{e}$,
D\'aniel N\'ogr\'adi$^{cd}$,
\speaker{Attila P\'asztor}$^{cd}$,
Andreas Sch\"afer$^e$,
K\'alm\'an K. Szab\'o$^{ab}$,
B\'alint C. T\'oth$^{a}$,
Norbert Trombit\'as$^{cd}$
\\
$^{a}$University of Wuppertal, Department of Physics, Wuppertal D-42097, Germany\\
$^{b}$J\"ulich Supercomputing Center, J\"ulich D-52425, Germany\\
$^{c}$E\"otv\"os University, Budapest 1117, Hungary\\
$^{d}$MTA-ELTE Lend\"ulet Lattice Gauge Theory Research Group\\
$^{e}$University of Regensburg, Regensburg D-93053, Germany
}

\abstract{
Finite temperature charmonium spectral functions in the pseudoscalar(PS) and vector(V) 
channels are studied in lattice QCD with 2+1 flavours of dynamical Wilson quarks, on fine 
isotropic lattices (with a lattice spacing of $0.057\rm{fm}$), 
with a non-physical pion mass of $m_{\pi} \approx 545 \rm{MeV}$. 
The highest temperature studied is
approximately 1.4$T_c$. Up to this temperature no significant variation of the
spectral function is seen in the PS channel. The V channel shows
some temperature dependence, which seems to be consistent with a temperature dependent
low frequency peak related to heavy quark transport, plus a 
temperature independent term at $\omega>0$. These results are in accord with previous 
calculations using the quenched approximation. The details of the study can be found in
\cite{sajat}.
}

\FullConference{ The 32nd International Symposium on Lattice Field Theory - Lattice 2014\\
June 23-28, 2014\\
Columbia University, New York, New York}

\begin{document}

\maketitle

\section{Introduction}

\subsection{Mesonic spectral functions}

The charmonium systems at finite T have been under heavy investigation, since the suppression of $J/\Psi$
is regarded as an important experimental signal for the formation of the plasma 
state \cite{CharmoniumOriginalPhenom}. Charmonium states are expected to dissolve somewhat
above the transition temperature, because of the screening of the inter-quark potential and
collisions with the hot medium. In this paper, we investigate the in medium properties of 
the $J/\Psi$ and $\eta_c$ mesons from a lattice QCD perspective. \\
The spectral function (SF) of a correlator of self-adjoint operators is the imaginary part of the Fourier-transform of the 
real time retarded correlator. In this article, we will deal with correlators between local charmonium meson currents, and 
the corresponding SFs. It can be shown, that the SF is related to the Euclidean correlator by an integral transform
\begin{equation}
\label{eq:IntegralTransform}
G(\tau,\vec{p}) = \int_0^\infty d\omega A(\omega,\vec{p}) K(\omega,\tau) \rm{,}
\end{equation}
where 
\begin{equation}
K(\omega,\tau) = \frac{\cosh(\omega(\tau-1/{2T}))}{\sinh(\omega/{2T})}
\end{equation}
is the integral kernel, and $G(\tau,\vec{p})$ is the Euclidean correlator (at zero chemical potential).
Knowledge of the SFs is of great importance. In the SF a stable particle gives a $\delta$ like peak, while an unstable particle in matter
gives a smeared peak. Also, the Kubo-formula states that the heavy quark diffusion constant D is related to the 
V spectral function as $D=\frac{1}{6 \chi} \lim_{\omega \to 0} \sum_{i=1}^3 \frac{A_{ii}(\omega,T)}{\omega}$ , 
where $\chi$ is the (heavy) quark number susceptibility and $A_{ii}$ is the spectral function corresponding to the V channel. 
If the transport coefficient is non vanishing, we expect some finite value of $\rho/\omega$ for small $\omega$. This implies the presence of a transport peak.
We will investigate the anticipated melting of the heavy meson states $J/\Psi$ and $\eta_c$ in the
quark gluon plasma. 

\subsection{The Maximum Entropy Method}
To get the SFs from a lattice study one has to invert equation (\ref{eq:IntegralTransform}). This 
inversion however is ill-defined, since the number of frequencies for which one
wants to reconstruct the SF is higher 
than the number of data points. In this case a $\chi^2$ fit on the shape of the SF discretized
to $N_\omega$ points is degenerate. One has to regularize the problem in some way. 
\footnote{Note that the number
of data points vs. parameters is not the only problem when trying to reconstruct SFs. Even if one has a large number
of data points (e.g. on time anisotropic lattices), the Euclidean correlator is rather insensitive to fine details of
the SF. Therefore the inversion introduces large uncertainties \cite{Aarts:2002cc}.}
The determination of hadronic SFs via the Maximum Entropy Method (MEM) was 
first suggested in \cite{Asakawa}. 
Here, one has to maximize:
\begin{equation}
\label{eq:Q}
Q = \alpha S - \frac{1}{2} \chi^2  \rm{,}
\end{equation}
where as usual
\begin{equation}
\chi^2 = \sum_{i,j=1}^{N_{\rm{data}}} (G_i^{\rm{fit}}-G_i^{\rm{data}}) C^{-1}_{ij} (G_j^{\rm{fit}}-G_j^{\rm{data}}) \rm{,}
\end{equation}
with $C_{ij}$ being the covariance matrix of the data (in Euclidean time), and the Shannon--Jaynes entropy is
\begin{equation}
S = \int d\omega \left( A(\omega)-m(\omega)-A(\omega) \log\left(\frac{A(\omega)}{m(\omega)}\right) \right) \rm{,}
\end{equation}
where the so called prior function $m(\omega)$ is supposed to summarize our prior knowledge on the shape of the 
SF (such as the leading perturbation theory behaviour).  After equation (\ref{eq:Q}) is maximized at a given value of $\alpha$, and the optimal $A_\alpha$ is obtained, 
the regularization parameter $\alpha$ has to be averaged over, with the conditional probability from \cite{Asakawa}. It can then be shown that the maximum of $Q$ at a given $\alpha$ lies in an $N_{\rm{data}}$ dimensional subspace of
the $N_\omega$ dimensional space of possible $A(\omega_i)$ vectors, that can be parametrized as 
$A(\omega)=m(\omega) \exp\left( \sum_{i=1}^{N_{\rm{data}}} s_i f_i(\omega)\right)$.
The particular parametrization of the subspace we use  is $f_i(\omega)=K(\omega,\tau_i)$ and was introduced 
by Ref. \cite{Jakovac}. In our experience this proved to be numerically more stable 
than the more widely known Bryan method\cite{Bryan}. The shape of the subspace is of course strongly dependent on the
choice of the prior function. This is the source of a systematic uncertainty, that has to be considered. 
We don't carry out an error analysis of the full spectral function, since with the current 
statistics that would give huge errors. Instead, we only give errors to some physically interesting 
quantities related to the spectral function, that are more stable. Systematic errors
have been estimated with varying the prior function shapes: $m_0 / a^2$, $m_0 \omega /a $, $m_0 \omega^2$, $a^{-3}/(m_0/a + \omega)$ and
the constant $m_0=0.001\rm{,}0.01\rm{,}0.1\rm{,}1.0$. Stastical errors have been estimated with 10 jackknife samples. We use a 
modified version of the kernel for the reconstruction \cite{Aarts:2007wj, Engels09}:
$\hat{K}\left(\tau,\omega\right)= \tanh \left( \omega/2 \right) K \left( \tau, \omega \right)$.
This ``cures" the low frequency divergent $1/\omega$ behaviour of the kernel, without spoiling the high $\omega$
behaviour. We also mention here, that the reliable use of the MEM, even with this method, which does not contain a 
SVD, still required arbitrary precision arithmetics in the implementation.

Lattice studies of charmonium SFs using the MEM have been carried out on numerous occasions, but 
so far not in 2+1 flavour QCD. A recent, detailed study of charmonium SFs in 
quenched QCD can be found in \cite{Ding}. Results regarding spectral functions 
with 2 flavours of dynamical quarks can be found in Ref. \cite{Kelly:2013cpa}. 

\subsection{Lattice configurations}

We use the same lattice configurations as in \cite{WilsonThermo}, details can be found there.
The gauge action used for the calculations was the Symanzik tree level
improved gauge action, and for the fermionic sector the clover improved 
Wilson action. Six steps of stout smearing with smearing parameter $\varrho=0.11$ were
used.  The clover coefficient was set to
its tree level value, $c_{\mathrm{SW}}=1.0$, which, for this type of smeared
fermions, essentially leads to an ${\cal O}(a)$ improved action. The full hadron spectrum
using this action was determined in Ref.~\cite{Science}. The bare masses of the $u$ and $d$ quarks were taken to be degenerate,
therefore the configurations were generated using an $N_f=2+1$ flavor
algorithm. The $u$ and $d$ quarks were implemented via the Hybrid Monte Carlo (HMC)
algorithm~\cite{Duane:1987de}, whereas the strange quark was implemented using
the Rational Hybrid Monte Carlo (RHMC) algorithm~\cite{Clark:2006fx}. 
From the study in \cite{WilsonThermo}, we only used the finest lattices, with
gauge coupling $\beta=3.85$, corresponding to a lattice spacing of $a=0.057(1)\rm{fm}$.
The bare light quark masses where choosen to be $am_{ud}=-0.00336$ and $am_s=0.0050$, which,
when fixing the scale with a physical $\Omega$ baryon mass, corresponds to a pion mass
$m_{\pi} = 545 \rm{MeV}$. The lattices had spatial extent $N_s^3=64^3$ and temporal
extent $N_t=28,20,18,16,14,12$. $N_t=28$ corresponds to $T=123$MeV, $N_t=12$ to $T=288$MeV.
As for the charm mass tuning, from ref. \cite{CharmMass} the ratio $m_c/m_s = 11.85$. Since with Wilson fermions, there 
is an additive renormalization, it is not possible to use this ratio directly in setting the
charm mass. However, we know that for $ud$ and $s$ the masses used in the simulation correspond to 
a mass ratio of 1.5 \cite{Durr:2010vn}, from this we get $(m_c-m_s)/(m_s-m_{ud})=35.55$ which gives the estimate
for the charm mass that was used. To check if this is approximately the correct charm mass,
we checked the masses of the different mesons states containing s and c quarks, and they were indeed
in the right ballpark. 

\section{Results}

\subsection{Zero temperature analysis: stability test}
Since the temperature is $T=1/(N_t a)$, as the temperature increases we have less 
and less data points for our reconstruction of the SFs. That means
that the reliability of the method decreases with increasing temperature. To the highest temperature, where the MEM results are still likely to be trusted,
we drop points (starting from $t=0$) from the lowest temperature correlators
and do a MEM reconstruction with these limited number of points. We say that the
reconstruction is no longer reliable when we can not reconstruct the first peak.
The result is that $N_t=14$ seams reliable still, but $N_t=12$ does not. 

\begin{figure}[t!]
\begin{center}
\vspace{-1cm}
\includegraphics[angle=0,width=0.495\linewidth]{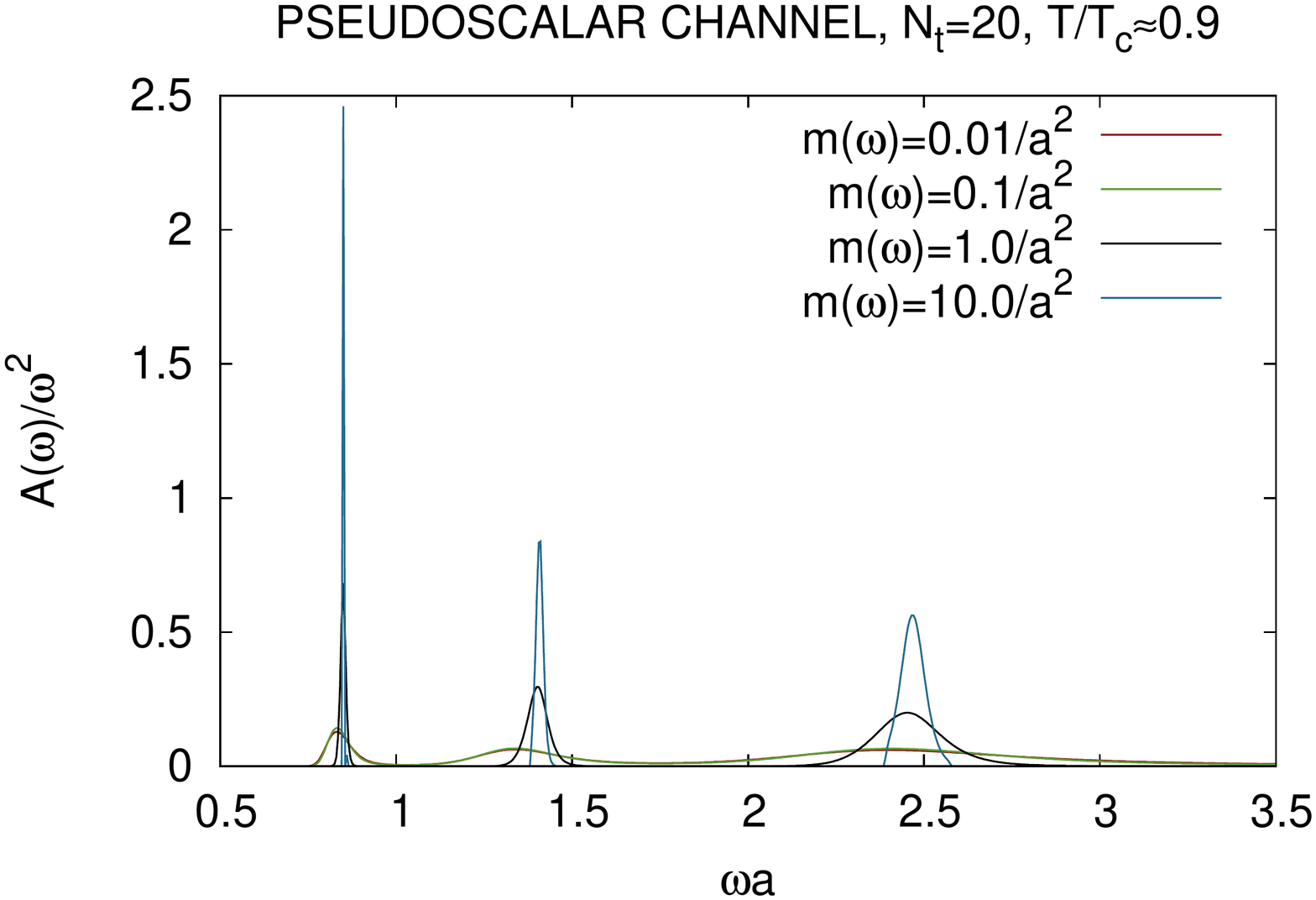}
\includegraphics[angle=0,width=0.495\linewidth]{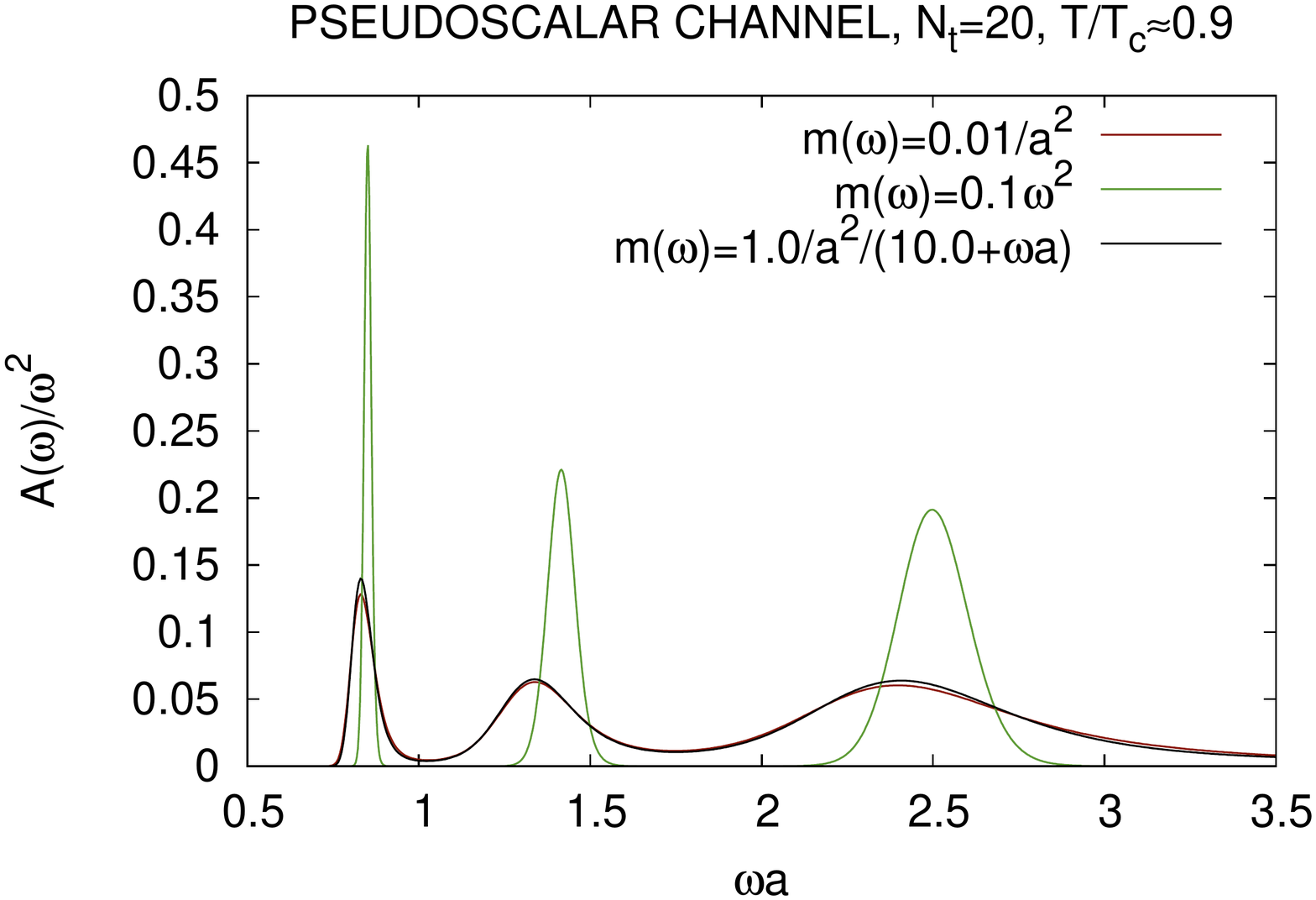} \\
\vspace{-1cm}
\includegraphics[angle=0,width=0.495\linewidth]{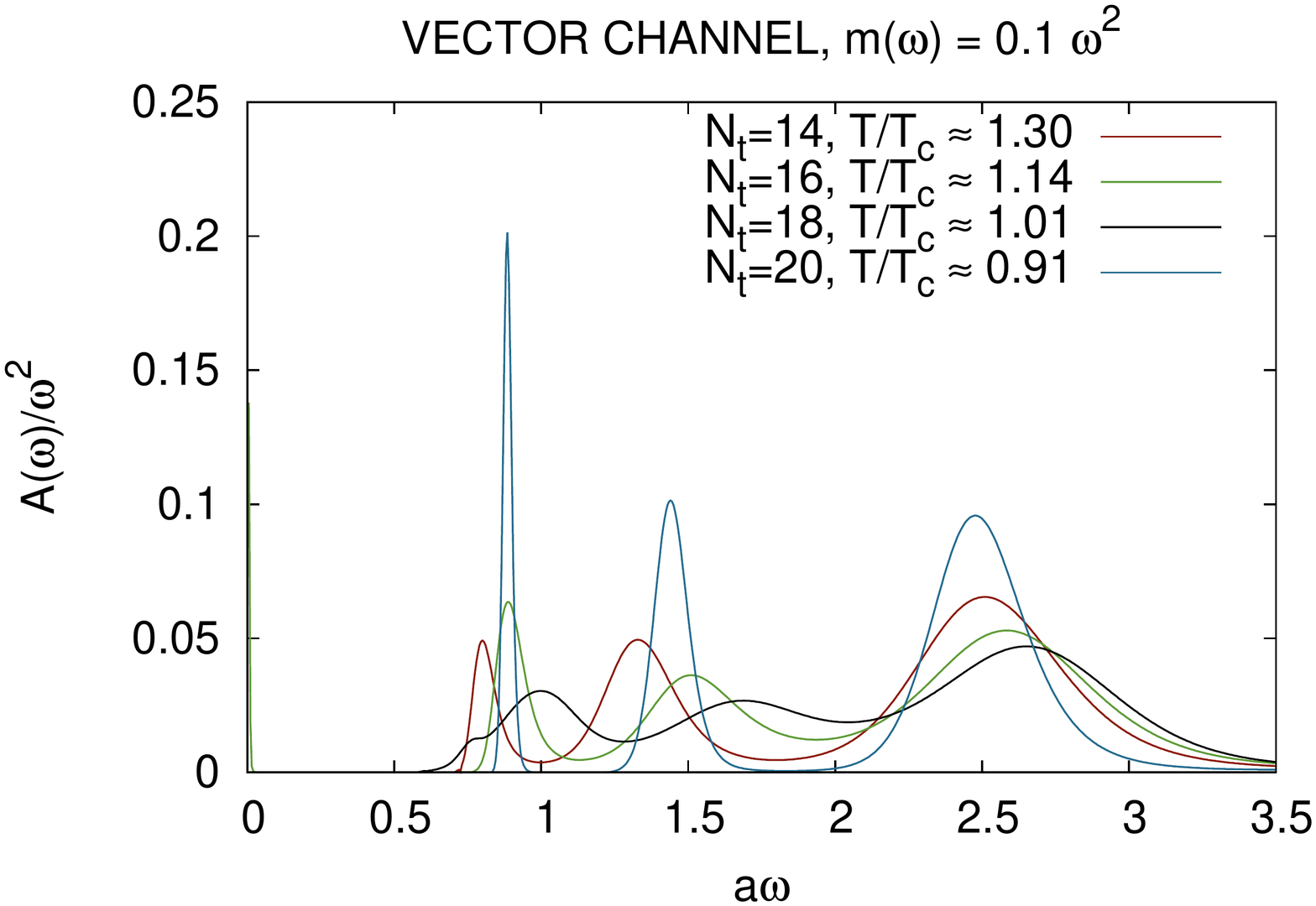} 
\includegraphics[angle=0,width=0.495\linewidth]{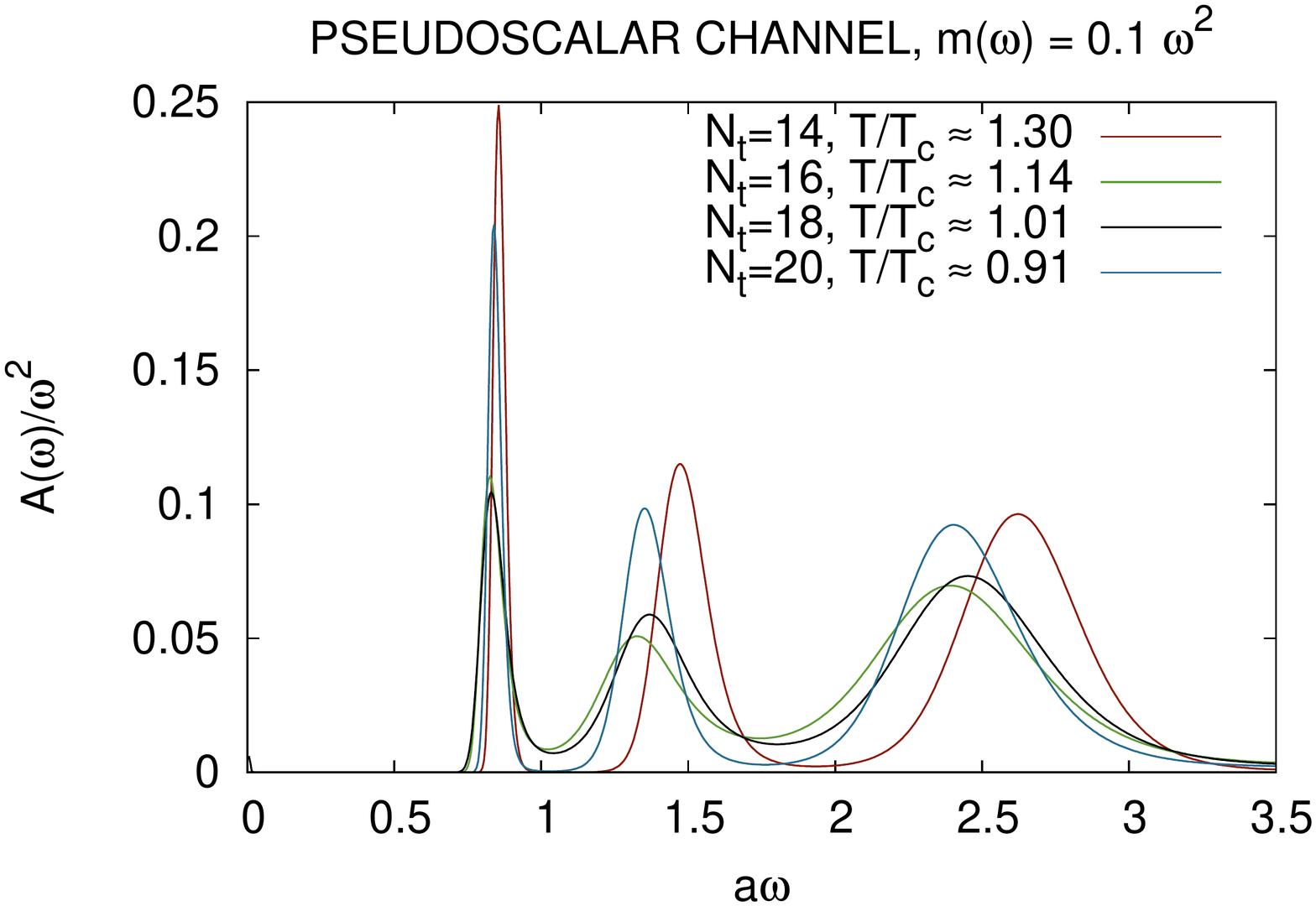} \\
\vspace{-1cm}
\includegraphics[angle=0,width=0.495\linewidth]{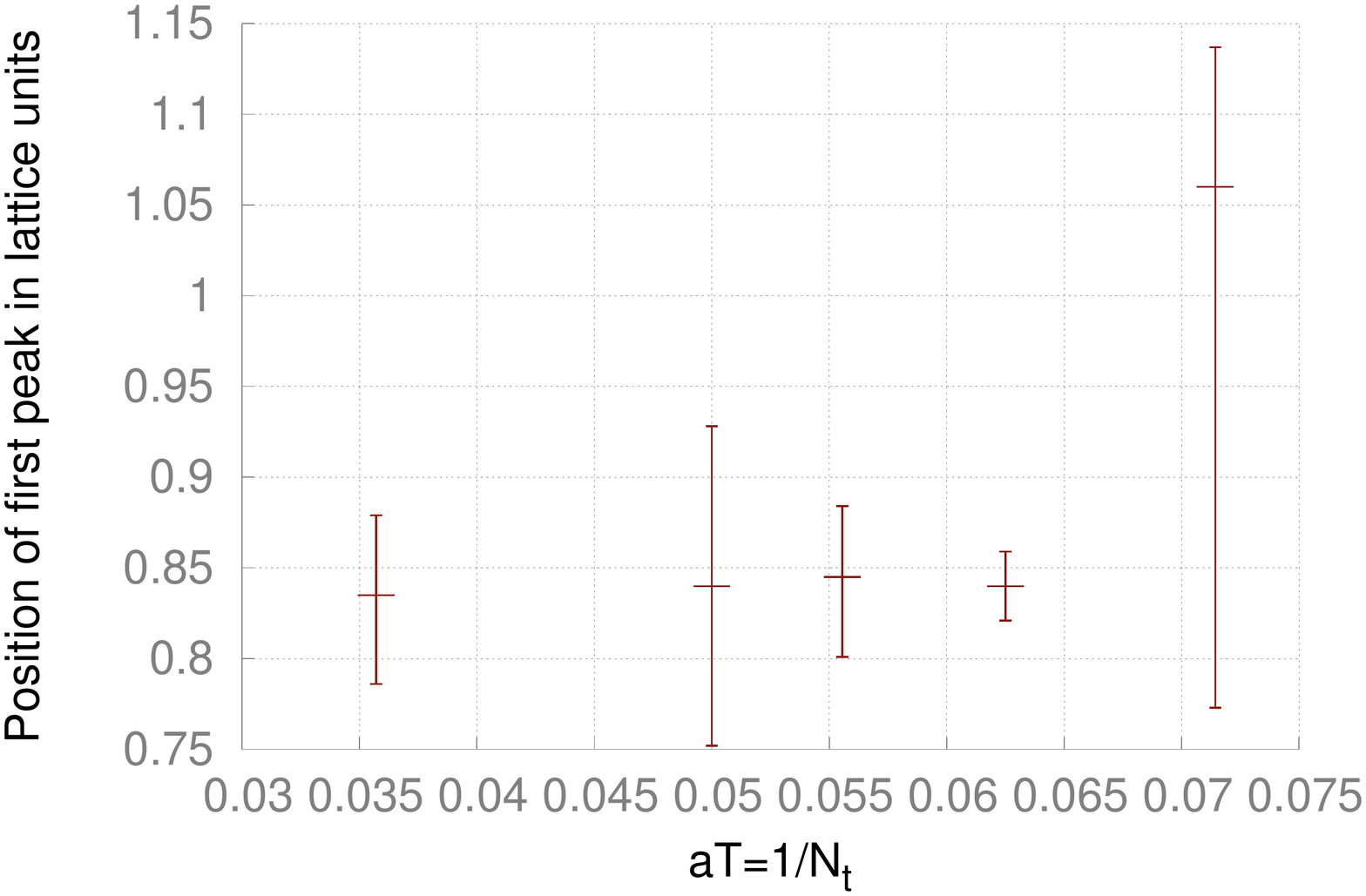}
\includegraphics[angle=0,width=0.495\linewidth]{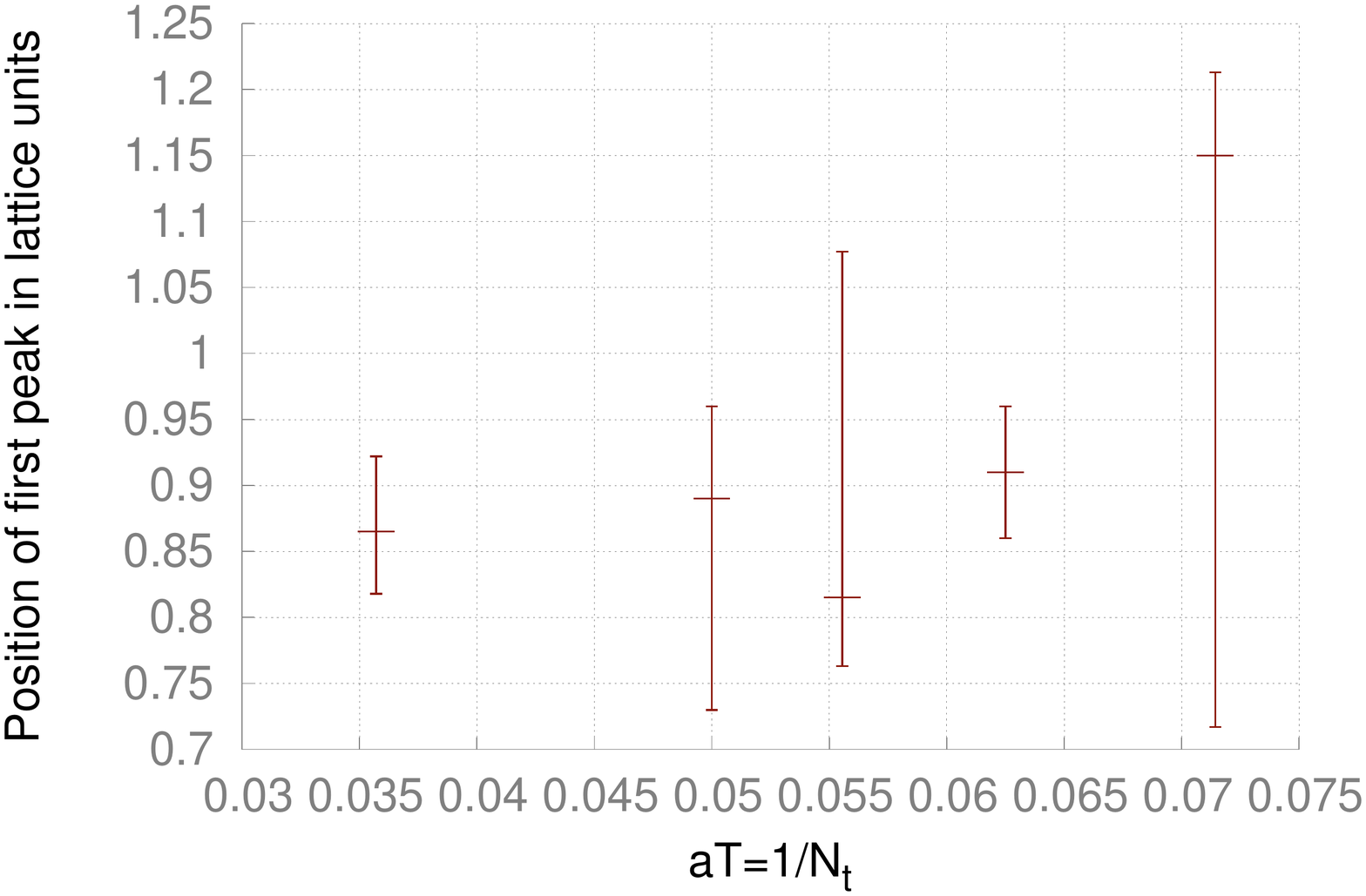}
\end{center}

\renewcommand{\figure}{Fig.}
\vspace{-1.2cm}
\caption{ The sensitivity of the reconstruction on the prior function(top).  
Only PS channel is included in the figure, but the V channel looks similar. 
The temperature dependence of the reconstucted spectral functions(middle).
The position of the first PS(left) and V(right) peak as a function of temperature(bottom).
}
\vspace{-0.3cm}
\label{fig:dirmem}
\end{figure}

\subsection{MEM reconstructed spectral functions}
So as far as our direct MEM analysis can tell, the PS SF is temperature
independent in the given range. See Figures \ref{fig:dirmem}.
The situation is a little bit more complicated in the V channel. Here, reconstruction of the SF 
shows some temperature dependence. The first peak appears to go down to lower energies at the highest temperature. 
Due to some properties of the analysis (i.e.
possible merging of adjacent peaks and problems with the resolution of the transport
peak) using MEM alone one cannot draw any firm conclusions about the
nature of the change in the SF - at least at the current level of statistical errors. 

\subsection{The ratio $G/G_{\rm{rec}}$}
\begin{figure}
\begin{center}
\includegraphics[angle=0,width=0.495\linewidth]{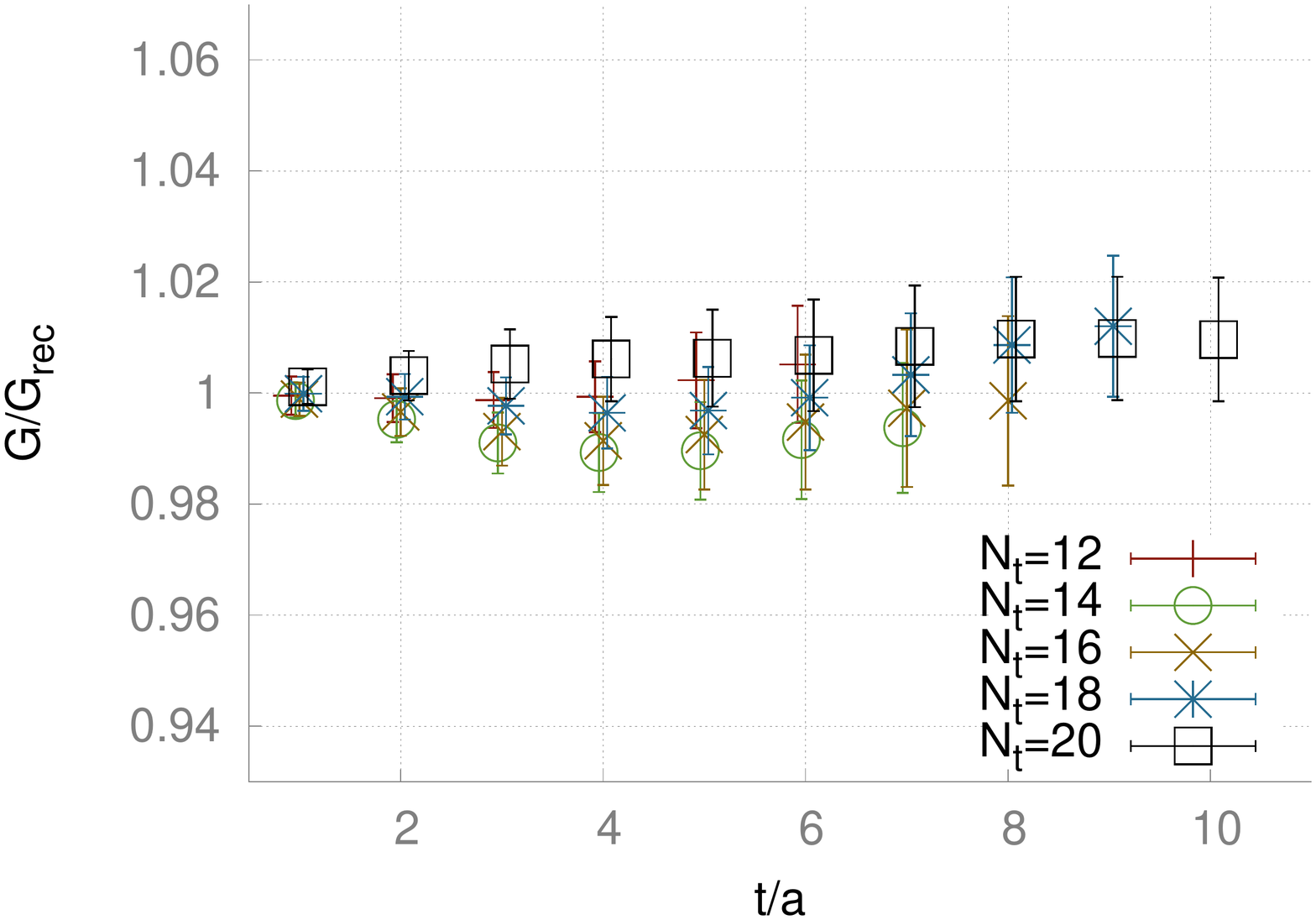} 
\includegraphics[angle=0,width=0.495\linewidth]{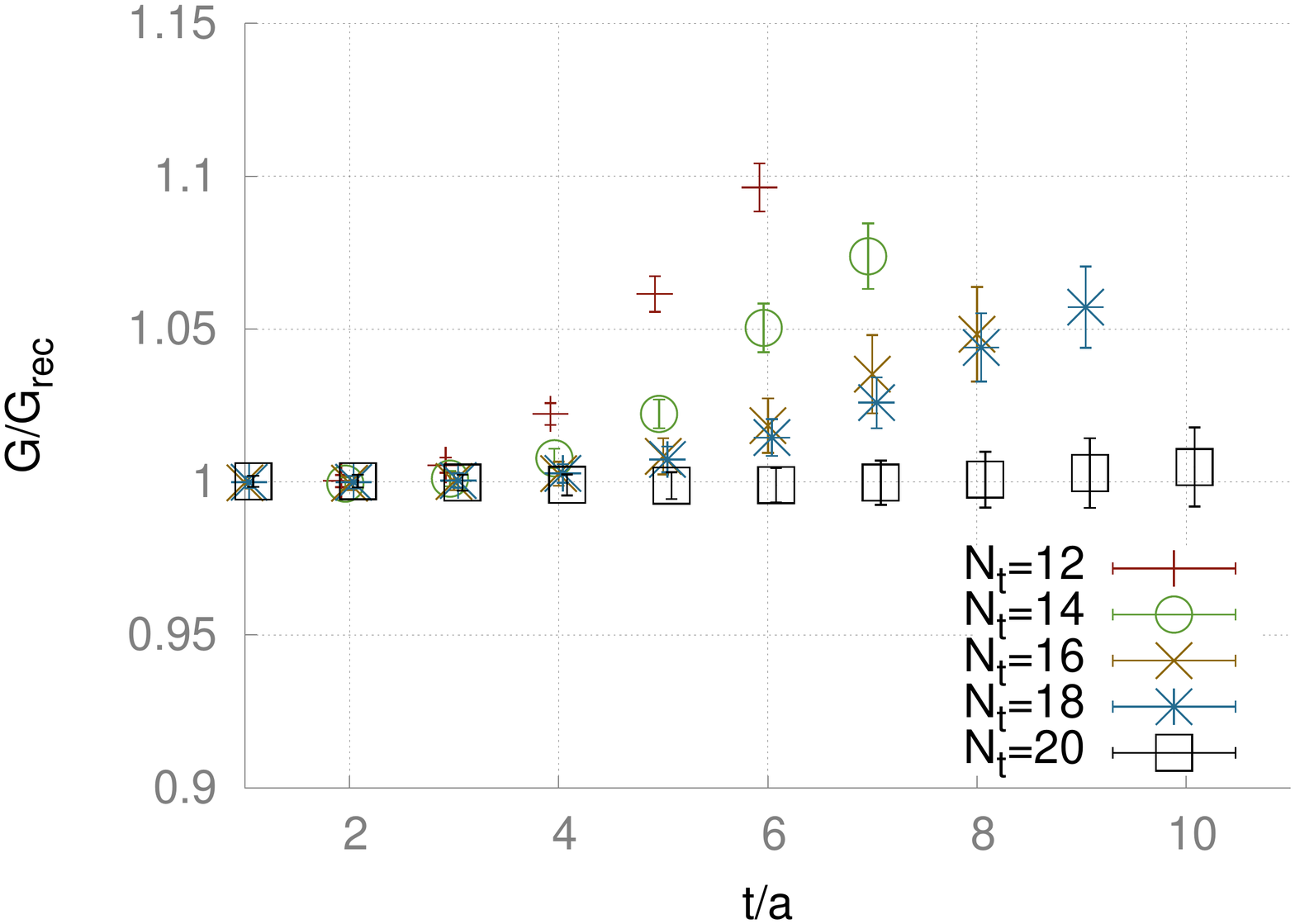} \\
\vspace{-1cm}
\includegraphics[angle=0,width=0.495\linewidth]{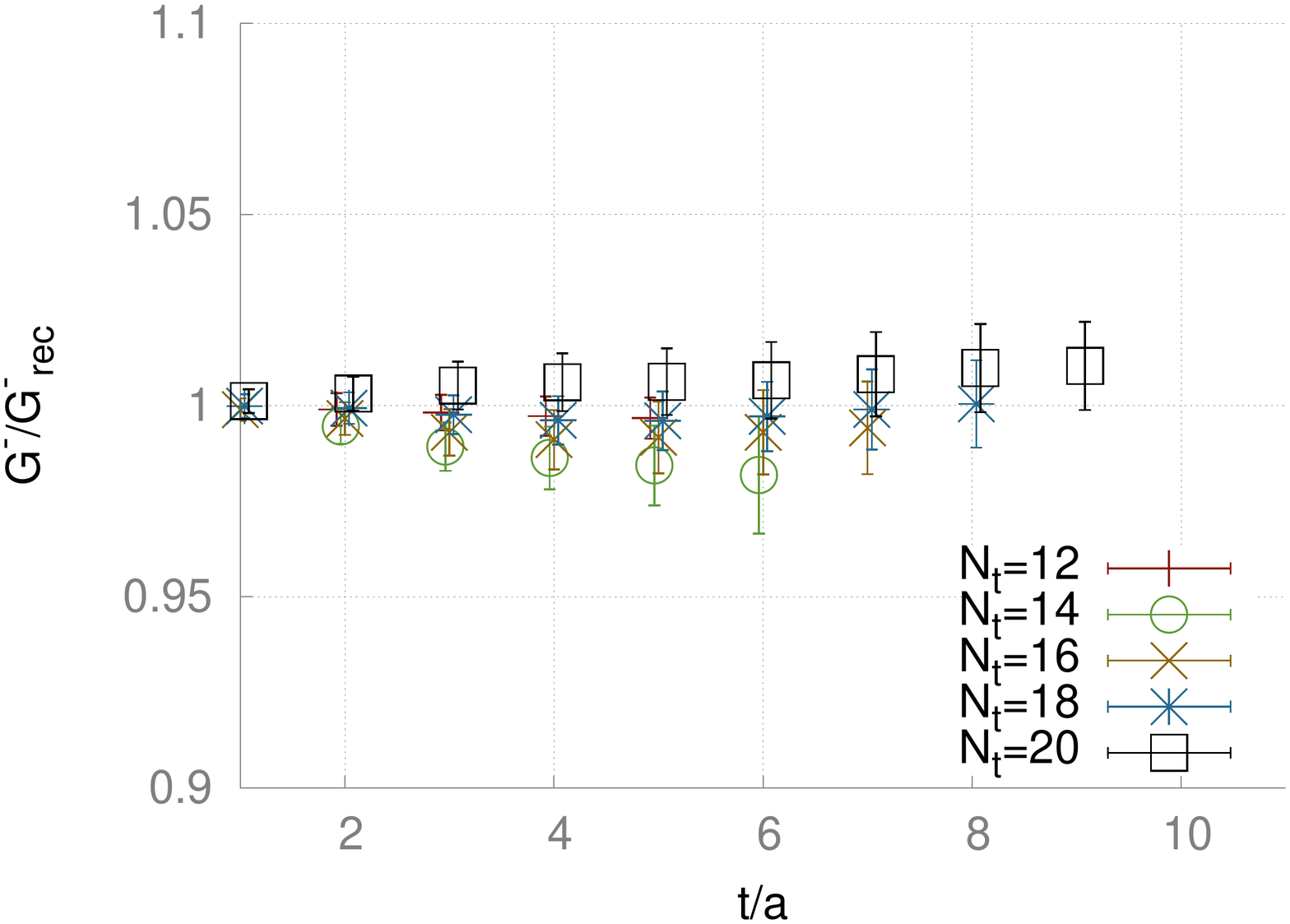} 
\includegraphics[angle=0,width=0.495\linewidth]{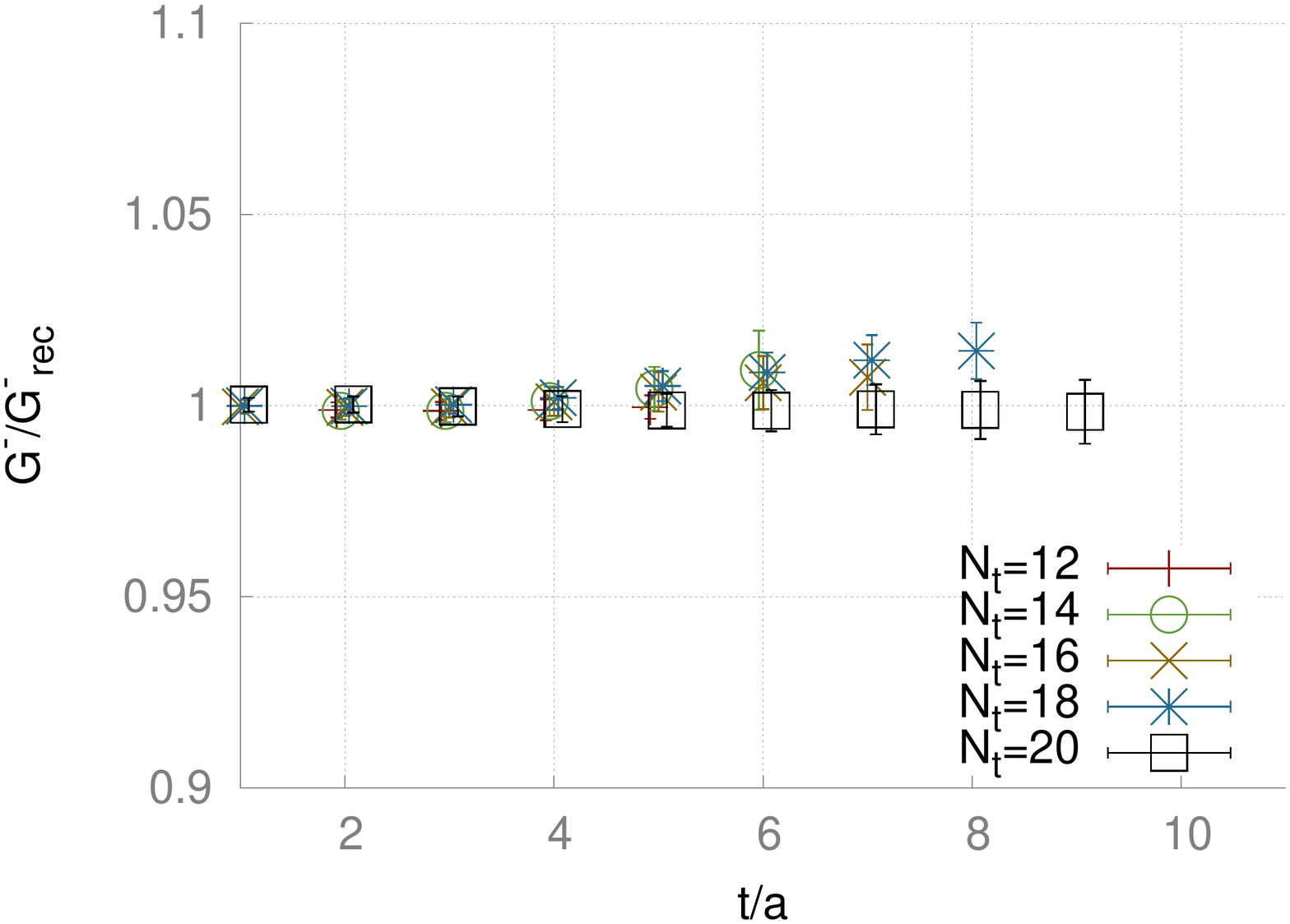}
\vspace{-1.2cm}
\caption{The ratio $G/G_{\rm{rec}}$ in the PS(left) and V(right) channels(). Also, the ratio $G^{-}/G^{-}_{\rm{rec}}$(bottom).}
\vspace{-1cm}
\label{fig:ratio}
\end{center}
\end{figure}

\begin{figure}[t!]
\begin{center}
\centering
\includegraphics[angle=0,width=0.495\linewidth]{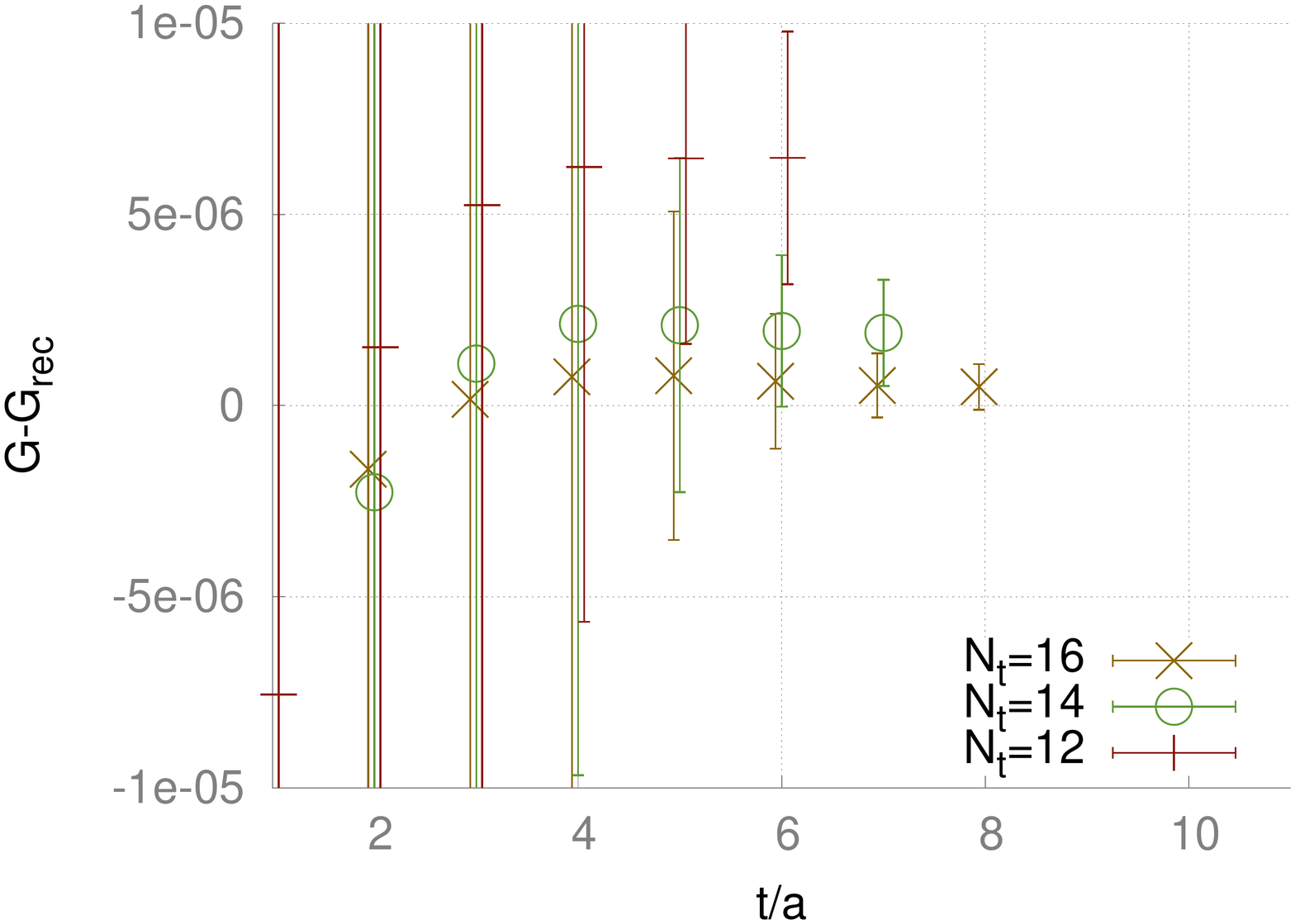}
\renewcommand{\figure}{Fig.}
\vspace{-0.7cm}
\caption{
The difference $G-G_{\rm{rec}}$ in the V channel at the 3 highest temperatures.}
\vspace{-1cm}
\label{fig:G_min_Grec}
\end{center}
\end{figure}

An alternative aproach to study spectral functions was suggested in \cite{Jakovac}. The ratio:
\begin{equation}
\frac{G \left( t, T \right) }{G_{\rm{rec}} \left( t, T \right)}  = \frac{ G(t,T) }{ \int A(\omega,T_{\rm{ref}})K(\omega,t,T)\rm{d}\omega }
\end{equation}
has a few advantages: MEM reconstruction is only needed at $T_{\rm{ref}}$, where we have the most data points, and 
so a more reliable reconstruction. We use $N_t=28$ as reference temperature.  We can calculate this ratio even at high temperatures, where the MEM reconstruction is
 already unreliable.  Also, if the spectral function is temperature independent, then the 
 trivial temperature dependence of the correlators, coming from the 
 integral kernel will drop out, and the ratio will be $G/G_{\rm{rec}}=1$. It is also useful to study the same ratio with mid-point subtracted correlators \cite{Umeda:2007hy}.
\begin{multline}
\frac{G^-}{G^-_{\rm{rec}}} = 
\frac{G \left( t, T \right) - G \left( N_t/2, T \right) }
{G_{\rm{rec}} \left( t, T \right) - G_{\rm{rec}} \left( N_t/2, T \right)}
= \frac{ G \left( t, T \right) - G \left( N_t/2, T \right) }
{ \int A(\omega,T_{\rm{ref}})\left[ K(\omega,t,T)-K(\omega,N_t/2,T) \right] \rm{d}\omega }
\end{multline}
This way, one can drop the zero-mode (constant) contribution to the correlators. These
have to do with transport coefficients, or other low frequency ($\omega \ll T$) features of the spectral
functions.  If the ratio of $G/G_{\rm{rec}}$ is
different from one, but the ratio with the middle-point substracted correlators is not, that
means that the temperature dependence of the SFs should be well described by just a zero-mode
contribution $f(T) \cdot \omega \delta(\omega-0^+)$. The results of such an analysis can be seen
in Figs. \ref{fig:ratio}. As one can see, the results in the PS
channel are consistent with a temperature independent SF, while the results in the V channel
show a temperature dependent zero mode/low frequency contribution in the SF. We can also try to 
extract the zero mode contribution itself by considering the difference $G-G_{\rm{rec}}$. This is only
plotted in the V channel, in Fig. \ref{fig:G_min_Grec} (in the PS channel it is always 
consistent with zero). The difference has big errors, but on the two highest temperatures it is 
non zero within 1$\sigma$. At every temperature it is consistent with a time separation
independent constant. With the ansatz $A(\omega,T)=f(T)\omega \delta(\omega-0^+)+A(\omega,T_0)$
we get $f(T)T \approx (3 \pm 1.5)\cdot 10^{-5}$ at $1.4T_c$ in lattice units \footnote{Since we are not
using the conserved current on the lattice, but a local current, this will have a finite, lattice spacing dependent
renormalization constant of $\mathcal{O}(1)$. We neglect this fact, since we don't do a continuum limit, and the
renormalization is temperature independent. }. This ansatz, taken strictly, would imply a diverging diffusion
constant. However, the data do not restrict the shape of the transport peak, they are only
sensitive to the area. By using this ansatz, we do not mean to say that the diffusion constant diverges,
we simply extract that area of the transport peak. To get a diffusion constant additional information is needed.
(The width or the height of the peak, which is too narrow to resolve at this point.) 
The survival of $J/\Psi$ up to such high temperatures is consistent with previous results in 
quenched and 2 flavour QCD (see eg. \cite{Ding, Kelly:2013cpa}). \\

\section{Summary}

We have performed a lattice study of charmonium spectral functions
with 2+1 dynamical Wilson quarks. The MEM reconstruction of the spectral functions is
hampered by the limited number of data points at higher temperatures, so the highest
temperature we used for MEM reconstruction was approximately $1.3T_c$.
The PS spectral functions did not show any noticable temperature variation. The V spectral
functions showed a temperature variation, which, according to the analysis of the ratio 
$G/G_{\rm{rec}}$ is consistent with a temperature dependent zero mode, and a temperature
independent non-zero part in the SFs. In conclusion, we can say that we observed no 
melting of the $\eta_c$ and $J/\Psi$ mesons up to temperature of $1.4T_c$, and we observe
no variations in the spectral functions of $\eta_c$ whatsoever.\\

\section*{Acknowledgment}

Computations were carried out on GPU clusters
at the Universities of Wuppertal and Budapest as well as on supercomputers in
Forschungszentrum Juelich. This work was supported by the EU Framework Programme 7 grant (FP7/2007-2013)/ERC No 208740, 
by the Deutsche Forschungsgemeinschaft grants FO 502/2, SFB-TR 55 and by Hungarian Scientific Research 
Fund grant OTKA-NF-104034.. \\


\end{document}